\documentclass[useAMS,usenatbib,letterpaper,onecolumn]{mn2e} 
\usepackage[totalwidth=480pt,totalheight=680pt]{geometry}
\usepackage{graphicx,amssymb,amsbsy,amsmath}
\newcommand{\be}{\begin{equation}}
\newcommand{\ee}{\end{equation}}
\def\lta{\,\raise 0.3 ex\hbox{$ < $}\kern -0.75 em
 \lower 0.7 ex\hbox{$\sim$}\,}
\def\gta{\,\raise 0.3 ex\hbox{$ > $}\kern -0.75 em
 \lower 0.7 ex\hbox{$\sim$}\,}

\newcommand{\mtot}{M_{\scriptstyle T} }

\newcommand{\selfg}{\alpha_{\rm g}}

\newcommand{\aphys}{a_0} 
\newcommand{\mrs}{H} 
\newcommand{\rof}{{\scriptstyle(f)}} 
\newcommand{\rofone}{{\scriptstyle(1-f)}} 
\newcommand{\vvv}{\overrightarrow} 
\newcommand{\rstarms}{R_{\ast({\rm\scriptscriptstyle MS)}}} 

\title[Energy Optimization in Binary Systems]
{Energy Optimization in Binary Star Systems: \\ 
Explanation for Equal Mass Members in Close Orbits}  

\author[Adams, Batygin, Bloch]{Fred C. Adams$^{1,2}$, 
Konstantin Batygin$^3$, and Anthony M. Bloch$^4$\\  
\\ 
$^1$Physics Department, University of Michigan, Ann Arbor, MI 48109\\
$^2$Astronomy Department, University of Michigan, Ann Arbor, MI 48109\\
$^3$Division of Geological and Planetary Sciences, 
California Institute of Technology, Pasadena, CA 91125\\
$^4$Math Department, University of Michigan, Ann Arbor, MI 48109 }

\begin{document} 

\date{$\quad$ February 2020} 

\maketitle

\label{firstpage}

\begin{abstract} 
Observations indicate that members of close stellar binaries often
have mass ratios close to unity, while longer-period systems
exhibit a more uniform mass-ratio distribution.  This paper
provides a theoretical explanation for this finding by determining the
tidal equilibrium states for binary star systems --- subject to the
constraints of conservation of angular momentum and constant total
mass. This work generalizes previous treatments by including the mass
fraction as a variable in the optimization problem. The results show
that the lowest energy state accessible to the system corresponds to
equal mass stars on a circular orbit, where the stellar spin angular
velocities are both synchronized and aligned with the orbit. These
features are roughly consistent with observed properties of close
binary systems. We also find the conditions required for this minimum
energy state to exist: [1] The total angular momentum must exceed a
critical value, [2] the orbital angular momentum must be three times
greater than the total spin angular momentum, and [3] the semimajor
axis is bounded from above. The last condition implies that
sufficiently wide binaries are not optimized with equal mass stars,
where the limiting binary separation occurs near
$\aphys\approx16R_\ast$.
\end{abstract} 

\begin{keywords}
Stars: Binaries: General; Stars: Binaries: Spectroscopic; \\
Stars: Early-Type; Stars: Formation
\end{keywords}

\section{Introduction} 
\label{sec:intro} 

A substantial fraction of the stellar population resides in binary
systems, which can influence stellar properties and their long-term
evolution. Binaries with short periods are especially important in
this context, as they can experience mass transfer and affect the
properties of both core-collapse and type Ia supernovae
\citep{demarco}. These processes, in turn, determine the merger rates
for compact objects, including those that produce sources for
gravitational waves (e.g., \citealt{narayan,phinney,riles}; see also
\citealt{lombardi}). In order to understand these dynamical systems,
previous studies have found the lowest energy states accessible to
binaries, subject to conservation of angular momentum
\citep{counselman,hut1980}.  These optimized states correspond to
circular orbits, where the angular velocities of both stars are
synchronized with the orbit, with all three angular momentum vectors
pointing in the same direction. These properties are often observed in
close binary systems (e.g., \citealt{meibom2006,mazeh2008}; see
\citealt{shubook} for a textbook discussion). In addition, however,
close binaries tend to have mass ratios that are closer to unity than
the population as a whole (see below) and the origins of this trend 
remain elusive (note that previous calculations were carried out for
systems with fixed stellar masses). The goal of this present work is
to extend previous energy optimization calculations to include the
apportionment of mass between the stellar members. Introducing this
additional degree of freedom, we find that the critical point of the
system is realized when the stars have equal masses. The existence of
this tidal equilibrium state requires a minimum value for the total
angular momentum. Moreover, in order for the critical point to be a
true minimum of the energy, rather than a saddle point, the orbital
angular momentum must be at least three times larger than the spin
angular momentum. In addition, when the self-gravity of the stars is
included in the energy budget, binary systems with equal mass
correspond to energy minima only for sufficiently close orbits.

Although the observational landscape is complicated, members of close
binary systems show a tendency towards equal masses, i.e., their mass
ratios $q\equiv m_1/m_2$ are significantly closer to unity than would
be expected if the secondaries sampled the full mass distribution
(e.g., \citealt{tokovinin2000,halbwachs2003}). In other words,
sufficiently close binaries tend to be twins (\citealt{pinson,elbadry}; 
cf. \citealt{lucy2006}). For solar-type stars with short periods, the
mass ratio distribution tends to rise toward higher values
\citep{mazeh1992,duqmayor}. A more recent survey \citep{moestefano}
finds that about 30\% of close binaries (with $P<20$ days and
solar-type primaries) have mass ratios in the range $0.9\le{q}\le1$.
Similarly, \cite{raghavan} find that the distribution of mass ratios 
for short-period binaries with solar-type primaries has a well-defined
peak near $q\sim1$ (see also \citealt{duchene,simonobbie}).  An excess
of twins (equal mass stars) is also observed for young binaries
\citep{kounkel}. For early-type stars, about 40\% of the close binary
sample (with peak separation near $a\sim0.15$ AU) shows nearly equal
masses $m_1\approx{m_2}$ (\citealt{kobulfryer}; cf.
\citealt{kobul2014}).  A similar result holds for a separate study
where the primaries are A-type stars \citep{derosa}. The data thus
show a significant excess of nearly equal mass stars for close
binaries. Moreover, the aforementioned studies indicate that the
distribution of mass ratios is significantly different for wider
binaries. Undoubtedly, the interplay of physical processes that leads
to the distributions of mass ratios for different binary separations
is complex, and can only be understood through detailed calculations.
Instead of taking this route, this paper provides a partial
explanation for these observational trends by showing that the lowest
energy state accessible to close binary systems has equal mass stars.

The general problem of finding the lowest energy states of a physical
system, subject to constraints, has a long history in astrophysics
(dating back to \citealt{darwin1,darwin2}). Previous applications
include binary star systems with fixed angular momentum
\citep{counselman,hut1980}, and ellipsoidal figures of equilibrium for
both isolated stars and binaries \citep{chandrasekhar,lai1993,levine};
these latter studies focus on energy optimization with respect to
  the compressibility and oblateness of the stars, whereas this paper
  focuses on optimization of the mass ratios. Tidal equilibrium
states\footnote{Note that the term `tidal equilibrium state'
  refers to the configuration of lowest energy, which can be reached
  through any type of energy dissipation --- not necessarily tidal
  forcing.}  have also been found for planetary systems containing Hot
Jupiters \citep{levrard,ab2015}, and for hierarchical triple systems
\citep{ab2016}. This procedure has been generalized for applications
to multi-planet systems with fixed orbital spacing \citep{adams2019},
where the optimal state corresponds to planets with nearly equal
masses. However, this prediction of mass uniformity breaks down when
self-gravity of the planets is included and the total mass of the
system is sufficiently large \citep{papertwo}. Although calculations
of this type --- by design --- gloss over the machinery of the
dissipative effects at play, they provide a powerful means of
illuminating the general outlines of dynamical evolution, subject to
energy damping.

This paper is organized as follows. We formulate the optimization
problem in Section \ref{sec:model}. In this treatment, the system
energy is a function of ten dimensionless variables, subject to
conservation of angular momentum (which provides three constraints).
Section \ref{sec:firstvary} derives the tidal equilibrium state, which
corresponds to equal mass stars on a circular orbit, with the orbital
angular momentum and stellar angular momentum aligned.  In Section
\ref{sec:secondvary}, we consider the second variation and find the
conditions required for the critical point to be the minimum of the
energy.  Section \ref{sec:obs} then compares the predictions of this
calculation with current observational data.  The paper concludes
in Section \ref{sec:conclude} with a summary of our results and a
discussion of their implications. 

\section{Formulation of the Optimization Problem}  
\label{sec:model} 

For binary systems, the energy budget includes the orbital energy, the
stellar spins, and the self-gravity of the constituent stars. The
total system energy can thus be written in the form 
\be
{\cal E} = - {G m_1 m_2 \over 2\aphys}  
- \selfg {G m_1^2 \over R_1} - \selfg {G m_2^2 \over R_2} + 
{1 \over 2} I_1 \Omega_1^2 + {1\over2} I_2 \Omega_2^2 \,. 
\ee
Here, the stars have masses $m_k$, radii $R_k$, and moments of inertia
$I_k$, where the $\Omega_k$ are the magnitudes of the stellar angular
velocity vectors. The semimajor axis of the orbit is given by $\aphys$ 
and $\selfg$ is a dimensionless factor of order unity (the constant 
$\selfg$ depends on the internal structure and is assumed to be the 
same for both stars; see Appendix \ref{sec:genforms} for further
discussion).

The square of the magnitude of the orbital angular momentum 
is given by 
\be
h^2 = G {m_1^2 m_2^2 \over m_1 + m_2} \aphys (1-e^2) \,, 
\ee
where $e$ is the eccentricity of the binary orbit. 
The total angular momentum can be written in the form 
\be
\vvv{\cal L} = 
\vvv{h} + I_1 \vvv{\Omega}_1 + I_2 \vvv{\Omega}_2 \,. 
\ee
For the sake of definiteness, we define the coordinate system so 
that the total angular momentum vector $\vvv{\cal L}$ points in 
the ${\hat z}$ direction and the orbital angular momentum vector 
$\vvv{h}$ lies in the $x$-$z$ plane. The angular velocity vectors 
of the stars can in principle have components in all three directions. 

The individual stellar masses are allowed to vary, but 
their sum $\mtot$ is assumed to be constant, so that 
\be
m_1 + m_2 = \mtot = constant\,. 
\ee 
We define the mass fraction $f$ so that 
\be
f = {m_1 \over \mtot} \qquad {\rm and} \qquad 
1-f = {m_2\over \mtot}\,.
\ee
For the stars, we adopt the approximate mass-radius relation 
\be
R_k = R \left({m_k \over \mtot} \right)^{1/2} \,, 
\label{massradius} 
\ee 
where $R$ is a constant (e.g., see \citealt{prialnik}). The scale $R$
is the radius of a star with mass $\mtot$ at the time when the masses
of the two stars are being determined. As a result, both stars have
radii $R_k<R$.  Moreover, since the masses are determined during the
early phases of evolution, $R$ is expected to be larger than the main
sequence value. With this choice of mass-radius relation, we can write
the moments of inertia in the form 
\be
I_k = \eta \, m_k R_k^2 = \eta\,\mtot R^2 
\left({m_k \over \mtot} \right)^{2} \,, 
\ee
where $\eta$ is a dimensionless quantity of order unity 
and is assumed to be the same for both stars. Note that 
both $\alpha_g$ and $\eta$ are determined by the density 
distributions of the stars.

Next we define dimensionless quantities, 
\be
a = {\aphys \over R}\,, \qquad 
E = {{\cal E} R \over G \mtot^2 } \,, \qquad  
L = { {\cal L} \over \mtot(G\mtot R)^{1/2}}\,,  \qquad 
\omega_k^2 = {\Omega_k^2 R^3 \over G\mtot} \,. 
\label{dimensionless} 
\ee
The expression for the dimensionless energy then takes the form 
\be
E = - {1 \over 2a} f (1-f) - 
\selfg \left[ f^{3/2} + (1-f)^{3/2} \right] + {1\over2} \eta 
\left[ \omega_1^2 f^2 + \omega_2^2 (1-f)^2 \right] \,,
\label{energy} 
\ee
where the dimensionless angular velocities each have three 
components so that 
\be
\omega_1^2 = \omega_{1x}^2 + \omega_{1y}^2 + \omega_{1z}^2 
\qquad {\rm and} \qquad 
\omega_2^2 = \omega_{2x}^2 + \omega_{2y}^2 + \omega_{2z}^2 \,. 
\ee
The components of the dimensionless angular momentum become 
\be
L_z = L = \sqrt{a} f(1-f) \cos i \sqrt{1-e^2} 
+ \eta f^2 \omega_{1z} + \eta (1-f)^2 \omega_{2z} \,,
\label{zangmom} 
\ee 
\be
L_x = 0 = \sqrt{a} f(1-f) \sin i \sqrt{1-e^2} 
+ \eta f^2 \omega_{1x} + \eta (1-f)^2 \omega_{2x} \,,
\label{xangmom} 
\ee
and finally 
\be
L_y = 0 = \eta f^2 \omega_{1y} + \eta (1-f)^2 \omega_{2y} \,. 
\label{yangmom} 
\ee 

The energy scale used in equation (\ref{dimensionless}) is comparable
to the binding energy of the stars, and $a\gta1$, so the dimensionless
energy $E$ is generally less than unity. The angular momentum scale is
comparable to that of a contact binary with total mass $\mtot$, so
that the dimensionless angular momentum $L$ is generally greater than
unity. The dimensionless stellar rotation rates $\omega_k$ are scaled
to the break-up speeds for a star of mass and radius $(\mtot,R)$. 
As a result, the stellar spins $\omega_k<1$, the spin angular momenta
$S_k\sim\eta\omega_k<1$, so that most of the angular momentum
typically resides in the orbit.

\section{Equilibrium States: The First Variation} 
\label{sec:firstvary} 

We need to find the minimum energy state for the system defined
through equation (\ref{energy}), subject to the constraint of constant
angular momentum (given by equations [\ref{zangmom}-\ref{yangmom}]).
The energy and angular momentum depend on the ten variables, i.e., 
\be
E=E({\vvv X}) \,, \qquad {\vvv L} = {\vvv L}({\vvv X}) \,, 
\qquad {\vvv X} = (a,e,i,\omega_{1x},\omega_{1y},\omega_{1z}, 
\omega_{2x},\omega_{2y},\omega_{2z},f) \,. 
\ee
For each variable (denoted here as $\xi$), the optimization 
condition has the form 
\be
{\partial E \over \partial \xi} + \vvv{\lambda} \cdot 
{\partial \vvv{L} \over \partial \xi} = 0 \,,
\ee
where the vector $\vvv{\lambda}\equiv(\lambda_x,\lambda_y,\lambda_z)$ 
is the Lagrange multiplier. 

\noindent
For semimajor axis $a$, we obtain the condition 
\be
{1 \over 2a^2} f(1-f) + {1 \over 2} a^{-1/2} f(1-f) \sqrt{1-e^2} 
\left[ \lambda_z \cos i + \lambda_x \sin i \right] = 0 \,.
\label{firsta} 
\ee
Note that solutions exist for the cases where $f=0$ or $f=1$, 
whereas we are interested in $f\ne0,1$. 

\noindent
Next we consider the eccentricity, which results in the condition 
\be
- {e \over \sqrt{1-e^2} } \sqrt{a} f(1-f) 
\left[ \lambda_z \cos i + \lambda_x \sin i \right] = 0 \,, 
\ee
and the inclination angle, which requires 
\be
\sqrt{a} f(1-f) \sqrt{1-e^2} 
\left[ - \lambda_z \sin i + \lambda_x \cos i \right] = 0 \,. 
\ee

\noindent
For components of the stellar rotation rate of the first 
star, all three equations have the form 
\be
\eta f^2 \left[ \omega_{1a} + \lambda_a \right] = 0 \,, 
\ee 
where the index $a$ runs over the three components $(a=x,y,z)$.
Similarly, for the stellar rotation rate of the second star we find 
\be
\eta (1-f)^2 \left[ \omega_{2a} + \lambda_a \right] = 0 \,. 
\ee 

\noindent
Finally, we consider the mass fraction $f$, which leads to 
the more complicated expression 
\be
- {1 \over 2a} (1-2f) - \selfg {3\over2} 
\left[ f^{1/2} - (1-f)^{1/2} \right] + \eta 
\left[ \omega_1^2 f - \omega_2^2 (1-f) \right] + 
\label{firstf} 
\ee
$$
\sqrt{a(1-e^2)} 
\left[ \lambda_z \cos i + \lambda_x \sin i \right] (1 - 2f) 
+ 2\eta f \vvv{\lambda} \cdot \vvv{\omega}_1 
- 2\eta (1-f) \vvv{\lambda} \cdot \vvv{\omega}_2 = 0 \,. 
$$

The solution to the first variation equations (\ref{firsta} --
\ref{firstf}), in conjunction with the definitions of the angular
momentum components (\ref{zangmom} -- \ref{yangmom}), yield the 
solution for the equilibrium point. The orbit must be circular 
and aligned with the stellar spins so that 
\be
\boxed{\qquad
e = i = \omega_{1x} = \omega_{2x} = \omega_{1y} = \omega_{2y} = 0 \,.
\qquad} 
\label{critzero} 
\ee
In addition, the stellar spin periods must match the orbital 
period, and the stars have equal mass, so that the nonzero 
system parameters have values 
\be
\boxed{\qquad
\omega_{1z} = \omega_{2z} = a^{-3/2}
\qquad {\rm and} \qquad f=1/2\,.
\qquad} 
\label{critpoint} 
\ee

The existence of the critical point implies a consistency constraint.
The angular momentum must be sufficiently large in order for the
critical point to exist. This result follows from evaluating the
angular momentum from equation (\ref{zangmom}) using the values
(\ref{critzero}, \ref{critpoint}) at the critical point, which leads 
to the expression 
\be
L = {1 \over 4} \left[ \omega^{-1/3} + 2\eta\omega\right] \,,
\ee
where we have suppressed the subscript on the stellar angular 
velocity $\omega$. This expression has a minimum value where  
\be
\omega = \left({1 \over 6\eta}\right)^{3/4} \,,
\ee
which implies the constraint 
\be
\boxed{\qquad
L > L_{\rm min} = {1\over3} 
\left[\,6 \eta \,\right]^{1/4} \,.  
\qquad} 
\label{lmin} 
\ee
If this condition is not met, then no critical point exists. In this
case, the binary system can attain a lower energy state by shrinking
the orbit --- and spinning up the stars to conserve angular momentum.
Note that the minimum $L$ constraint is evaluated at the critical 
point where $f=1/2$, corresponding to equal mass stars. In this case, 
the minimum angular momentum corresponds to an extremely close orbit 
(some type of contact binary) so that equation (\ref{lmin}) is 
almost always satisfied. In contrast, this minimum angular momentum
does provide non-trivial constraints for binary systems with extreme 
mass ratios (see \citealt{levrard,ab2015}).  

\begin{figure} 
\centerline{ \includegraphics[width=0.80\textwidth]{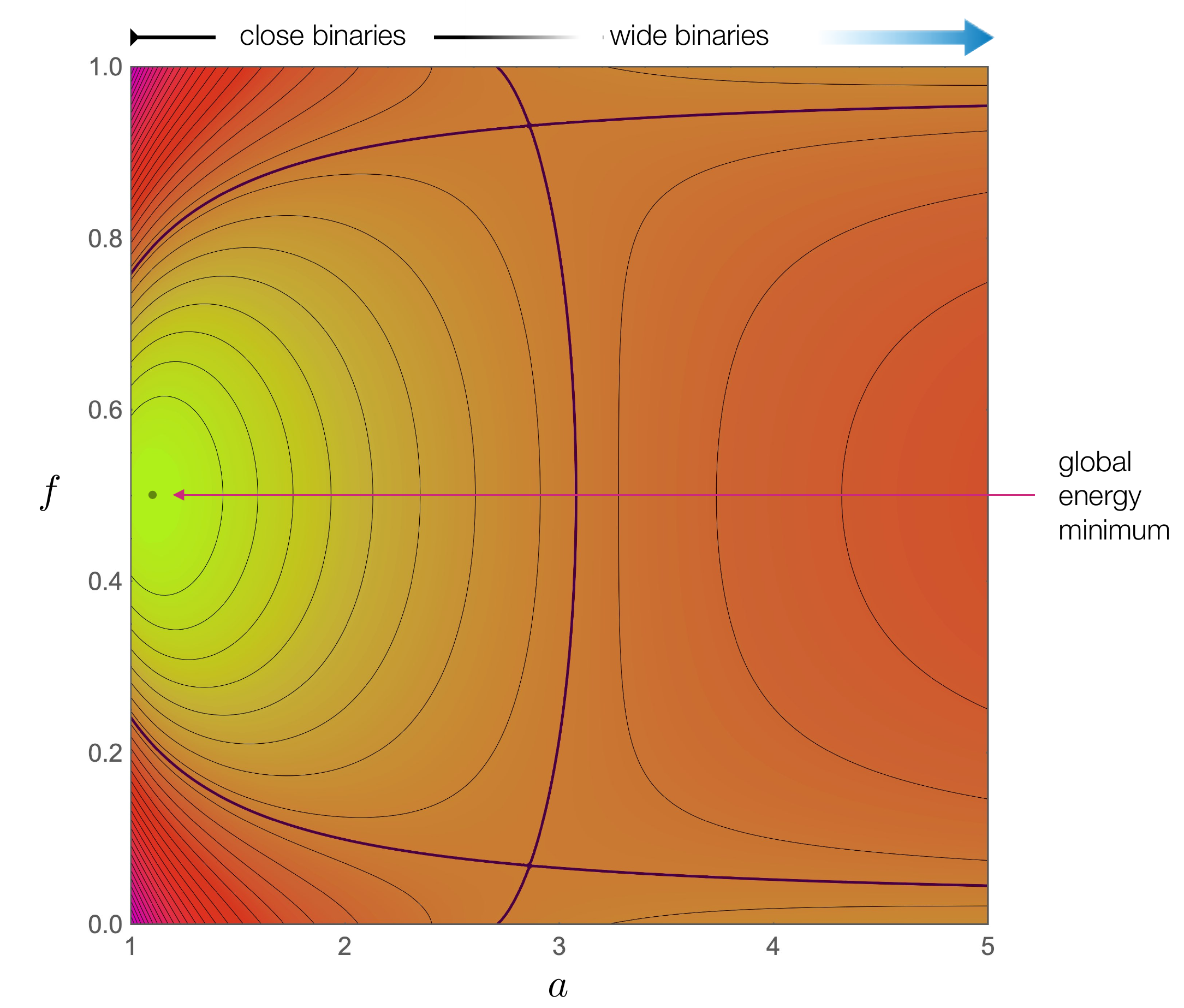} }
\caption{Contour plot showing the energy as a function of the 
mass fraction $f$ and the semimajor axis $a$ of the binary orbit 
(in dimensionless form). The other variables are taken to be 
those of the tidal equilibrium state, so that
$e=i=\omega_{1x}=\omega_{2x}=\omega_{1y}=\omega_{2y}=0$, and
$\omega_{1z}=\omega_{2z}=a^{-3/2}$. The region shaded yellow 
(centered on $f=1/2$ for small $a$) represents the lowest energy
states available to the system over this part of the plane, and
corresponds to equal mass stars. At larger values of $a$, however, the
energy has a maximum at $f=1/2$ (shaded orange on the right side of 
the diagram). In this region, lower energy states are available as
$f\to0,1$, favoring stars with unequal masses. The transition from an
energy minimum to an energy maximum occurs at $a\approx3$ (for this
choice of parameters -- see text). }
\label{fig:contour} 
\end{figure}  

Figure \ref{fig:contour} shows a contour plot of the energy as a
function of the mass fraction $f$ and the semimajor axis $a$ of the
binary orbit. In order to illustrate basic trends, the eccentricity,
inclination angle, and stellar spin components are specified to be
those of the tidal equilibrium state (given by equations
[\ref{critzero}, \ref{critpoint}]).  The stellar structure parameters
are taken to be $\eta=0.2$ and $\selfg=0.15$.  In this state, the
angular momentum $L$ = $\sqrt{a}/4$ + $\eta/(2a^{3/2})$, so the values
of $a$ in the diagram correspond to different choices of total angular
momentum. For small semimajor axis, $a\sim1-2$, Figure
\ref{fig:contour} displays a well defined minimum for nearly equal
mass stars with $f=1/2$ (shown as the yellow region). This energy
minimum for $f=1/2$ turns into a maximum for larger semimajor axes
(shown as the orange region on the right side of the diagram), 
with the transition near $a\gta3$. However, the strength of the
maximum at large $a$ (as measured by the energy difference between
$f=1/2$ and $f=0,1$) is much weaker than the depth of the energy
minimum at small $a$. This asymmetry implies that the tendency for
close binaries to have equal masses should be stronger than the
tendency for wider binaries (here $a\gta3$) to have small mass ratios.

Although Figure \ref{fig:contour} illustrates basic trends, many of
the independent variables have been suppressed. The following section
considers the full second variation of the energy function and thus
provides a more robust assessment of the conditions required for the
tidal equilibrium state to be a minimum energy state.

\section{Stability of the Critical State: The Second Variation} 
\label{sec:secondvary} 

In order to show that the critical point found in the previous section
corresponds to a stable equilibrium point, we must consider the second
variation. For this analysis, it is useful to explicitly conserve
angular momentum (instead of using Lagrange multipliers), so that we 
write the expression for the energy in the form  
\be
E = - {1 \over 2a} f (1-f) - 
\selfg \left[ f^{3/2} + (1-f)^{3/2} \right] \hskip2.0truecm
\ee
$$
\hskip2.0truecm + {1\over2} \eta (1-f)^2 
\left[ \omega_{2x}^2 + \omega_{2y}^2 + \omega_{2z}^2 \right]
+ {1 \over 2 \eta f^2} 
\left[ S_{1x}^2 + S_{1y}^2 + S_{1z}^2 \right] \,. 
$$
For shorthand notation, we have defined the components 
of the spin angular momentum of the first star as 
\be
S_{1z} \equiv \left[ L - \sqrt{a} f(1-f) \cos i \sqrt{1-e^2} 
- \eta (1-f)^2 \omega_{2z} \right] \,, 
\ee 
$$
S_{1x} \equiv - \left[ \sqrt{a} f(1-f) \sin i \sqrt{1-e^2} 
+ \eta (1-f)^2 \omega_{2x} \right]\,, \qquad {\rm and} \qquad 
S_{1y} \equiv - \eta (1-f)^2 \omega_{2y} \,.
$$
This formulation of the problem explicitly enforces 
conservation of angular momentum. 

We now take derivative with respect to all of the remaining 
variables. For the binary semimajor axis $a$: 
\be
E_a = {1 \over 2a^2} f (1-f) - {1 \over 2 \eta f^2} 
\left[ a^{-1/2} f(1-f) \sqrt{1-e^2} \right] 
\left[ S_{1x} \sin i + S_{1z} \cos i \right] \,. 
\ee 
For the eccentricity $e$: 
\be
E_e = {1 \over \eta f^2} 
\left[ S_{1x} \sin i + S_{1z} \cos i \right] 
\left[ \sqrt{a} f(1-f) \right] 
{e \over \sqrt{1-e^2} } \,. 
\ee
For the inclination angle $i$: 
\be
E_i = {1 \over \eta f^2} 
\left[ S_{1x} \cos i - S_{1z} \sin i \right] 
\left[ - \sqrt{a} f(1-f) \sqrt{1-e^2} \right] 
\ee
For the components of the stellar angular velocity 
$\omega_{2k}$, we get three expressions:  
\be 
E_{\omega2z} = \eta (1-f)^2 \omega_{2z} -
{1 \over f^2} \left[ S_{1z} (1-f)^2 \right] \,,
\ee 
\be
E_{\omega2x} = \eta (1-f)^2 \omega_{2x} - 
{1 \over f^2} \left[ S_{1x} (1-f)^2 \right] \,,
\ee
and 
\be
E_{\omega2y} = \eta (1-f)^2 \omega_{2y} - 
{1 \over f^2} \left[ S_{1y} (1-f)^2 \right] \,. 
\ee
And finally for the mass fraction $f$:  
\be 
E_f = - {1 \over 2a} (1-2f) - \selfg 
{3\over2} \left[ f^{1/2} - (1-f)^{1/2} \right] 
- \eta (1-f) \omega_2^2 
- {1 \over \eta f^3} 
\left[ S_{1x}^2 + S_{1y}^2 + S_{1z}^2 \right] 
\ee
$$
+ {1 \over \eta f^2} 
\left\{  S_{1x} \left[ - \sqrt{a} (1-2f) \sin i \sqrt{1-e^2} 
+ \eta 2(1-f) \omega_{2x} \right] \right\} 
$$
$$ 
+ {1 \over \eta f^2} \left\{  
S_{1y} \eta 2(1-f) \omega_{2y} + 
S_{1z} \left[ - \sqrt{a} (1-2f) \cos i \sqrt{1-e^2} 
+ \eta 2 (1-f) \omega_{2z} \right] \right\} \,.
$$  

\noindent
Next we find (again) the critical point. The solution corresponds to
$e=i=0$, $\omega_{1x}=\omega_{2x}=0$, $\omega_{1y}=\omega_{2y}=0$, and
$\omega_{1z}=\omega_{2z}=\omega\ne0$. With these specifications, the
remaining two equations can be evaluated.

\noindent
For the $E_a=0$ condition we obtain
\be
{1\over a^{3/2}} \eta f^2 = S_{1z} = L - h - \eta \omega (1-f)^2 \,. 
\ee 
For the $E_f=0$ condition we obtain 
\be 
E_f = - {1 \over 2a} (1-2f) - \selfg 
{3\over2} \left[ f^{1/2} - (1-f)^{1/2} \right] 
- \eta (1-f) \omega^2 
- {1 \over \eta f^3} 
\left[ S_{1z}^2 \right] 
\ee
$$ 
+ {1 \over \eta f^2} \left\{  
S_{1z} \left[ - \sqrt{a} (1-2f) 
+ \eta 2 (1-f) \omega \right] \right\} = 0 \,.
$$  
The first condition shows that $\omega=a^{-3/2}$ so that 
$S_{1z}=\eta \omega f^2$. Using these results, the final 
equation has solution $f=1/2$. We thus recover the same 
critical point found earlier, as expected (see equations 
[\ref{critzero}, \ref{critpoint}]).

Next we need to evaluate the Hessian matrix and find its eigenvalues,
or at least find the criteria necessary for the eigenvalues to be
positive \citep{hesse}. First consider the second derivatives with
respect to a single variable. For the semimajor axis matrix element,
we find the expression 
\be
E_{aa} = {1 \over 16a^3} \left[-3 + {a^2\over \eta}\right] \,. 
\ee
For the eccentricity matrix element, we get 
\be
E_{ee} = {1\over4} \omega\sqrt{a} = {1 \over 4a} \,. 
\ee
For the inclination matrix element, we get 
\be
E_{ii} = {1\over4} \omega\sqrt{a} + {a \over 4\eta} 
= {1 \over 4a} + {a \over 4\eta} \,. 
\ee
The second derivatives with respect to each of the 
components $\omega_{2a}$ all have the same form,  
\be
E_{\omega2x,\omega2x} = E_{\omega2y,\omega2y} = 
E_{\omega2z,\omega2z} = {\eta \over 2} \,. 
\ee
The second $f$ derivative has the form 
\be 
E_{f\!f} = {1 \over a} - \selfg {3\over4} 
\left[ f^{-1/2} + (1-f)^{-1/2} \right] + \eta \omega^2 
\ee
$$
+ (1/\eta f^2) \left[ - \sqrt{a} (1-2f) + \eta 2(1-f) \omega \right]^2 
+ (S_{1z}/\eta f^2) \left[ 2 \sqrt{a} - 2\eta \omega \right] 
$$
$$
- (4S_{1z}/\eta f^3) \left[ - \sqrt{a} (1-2f) 
+ \eta 2(1-f) \omega \right] + 3 S_{1z}^2/(\eta f^4) \,.
$$  
This expression can be evaluated at the critical point and 
simplified to find the matrix element 
\be 
E_{f\!f} = {3 \over a} - \selfg {3\over2} \sqrt{2} - {2 \eta \over a^3} \,. 
\label{secondff} 
\ee

Now we consider the mixed derivatives. For derivatives involving 
the semimajor axis, $E_{a\xi}$, the only nonzero term is
\be
E_{a\omega2z} = {1\over8 \sqrt{a}} \,.
\ee
For the case of eccentricity, $E_{e\xi}=0$ for all mixed 
derivatives. For inclination angle, $E_{i\xi}$, the only 
nonzero term is 
\be
E_{i\omega2x} = {\sqrt{a} \over 4} \,. 
\ee
All of the other terms are zero. 

Here we order the variables according to
($a,\omega_{2z},i,\omega_{2x},e,\omega_{2y},f$). The entire $7\times7$
Hessian matrix $\mathbb{H}$ takes the block diagonal form 
\be
\mathbb{H} = \left[
\begin{matrix} 
\mathbb{A} & 0 & 0 \\
0 & \mathbb{B} & 0 \\ 
0 & 0 & \mathbb{C} \\
\end{matrix} 
\right] \,,
\ee
where the submatrices are defined as follows: 
For the variables $(a,\omega_{2z})$, we obtain 
the $2\times2$ matrix  
\be
\mathbb{A} = \left[
\begin{matrix} 
(a^2 - 3\eta)/(16\eta a^3) & 1/8\sqrt{a} \\
\\
1/8\sqrt{a} & \eta/2 \\ 
\end{matrix} 
\right] \,. 
\ee
For the variables $(i,\omega_{2x})$, we find a 
second $2\times2$ matrix that takes the form 
\be
\mathbb{B} = \left[
\begin{matrix} 
1/4a + a/4\eta & \sqrt{a}/4 \\
\\
\sqrt{a}/4 & \eta/2 \\ 
\end{matrix} 
\right] \,.
\ee
For the remaining three variables $(e,\omega_{2y},f)$, 
all of the mixed derivatives vanish, so that we obtain 
the diagonal $3\times3$ matrix 
\be
\mathbb{C} = \left[
\begin{matrix} 
1/4a & 0 & 0 \\ 
\\
0 & \eta/2 & 0 \\ 
\\
0 & 0 & 3/a - \selfg 3 \sqrt{2}/2 - 2 \eta/a^3 \\ 
\end{matrix} 
\right] \,.  
\ee 

The eigenvalues of the $2\times2$ submatrix $\mathbb{A}$ for the
variables $(a,\omega_{2z})$ are determined by the solution to the
quadratic equation 
\be
\lambda^2 - [(a^2 - 3\eta)/(16\eta a^3) + \eta/2]\lambda + 
(a^2 - 3\eta)/(32a^3) - 1/64a = 0 \,,
\ee
which has solutions 
\be
2\lambda = \left[{a^2-3\eta \over 16\eta a^3} + {\eta\over2}\right]\pm
\left\{ \left[{a^2-3\eta \over 16\eta a^3} + {\eta\over2}\right]^2 - 
\left[{a^2-3\eta \over 8a^3} - {1\over16a}\right] \right\}^{1/2} \,.
\ee
Both eigenvalues will be positive provided that the final term 
in square brackets is positive, which enforces the condition 
\be
a^2 > 6\eta \,. 
\label{constraint} 
\ee
Since we do not need the eigenvalues themselves --- only their signs
--- this constraint can also be derived by showing that the matrix is
positive definite (so that it has only positive eigenvalues). From
Sylvester's Criterion, a real-symmetric matrix is positive definite 
if and only if all its leading principal minors are positive 
\citep{gilbert1991}. Applying this criterion to the matrix
$\mathbb{A}$ leads to the result (\ref{constraint}).

The interpretation of the constraint (\ref{constraint}) is that the
system can only enter into its equilibrium state if the orbital
angular momentum $h$ is sufficiently large compared to the total spin
angular momentum $S_T = I_1\omega_1 + I_2\omega_2$. In dimensionless
units, in the critical state, $h=\sqrt{a}/4$ and $S_T=(1/2)\eta
a^{-3/2}$, so the ratio $h/S_T=a^2/2\eta$. Equation (\ref{constraint})
thus implies that the orbital angular momentum must be at least 3
times greater than the total spin angular momentum in order for the
extremal state to be an energy minimum (and hence stable), i.e., 
\be
\boxed{\qquad h > 3S_T = 3(S_1+S_2) \,. \qquad}
\label{h3s} 
\ee

The eigenvalues of the $2\times2$ submatrix $\mathbb{B}$ for 
the variables $(i,\omega_{2x})$ are given by the quadratic
\be
\lambda^2 - \left({1\over4a}+{a\over4\eta}+{\eta\over2}\right) \lambda 
+ {\eta\over2} \left({1\over4a}+{a\over4\eta}\right) - {a\over16}=0\,,
\ee
which has solutions 
\be
2\lambda = \left({1\over4a}+{a\over4\eta}+{\eta\over2}\right) 
\pm \left\{ \left({1\over4a}+{a\over4\eta}+{\eta\over2}\right)^2 
- \left({\eta\over2a}+{a\over4}\right) \right\}^{1/2}\,. 
\ee
Both of these eigenvalues are always positive. 

The eigenvalues of the third $3\times3$ matrix $\mathbb{A}$ for the
variables $(e,\omega_{2y},f)$ are given by the diagonal elements. The
first two are manifestly positive. The final eigenvalue for the mass
fraction $f$ has the form 
\be
\lambda_f = {3\over a} - \selfg {3 \sqrt{2}\over2} 
- {2 \eta\over a^3} \,. 
\ee
Stability requires that $\lambda_f>0$, which is turn 
places a constraint on the system parameters 
\be
\selfg < {\sqrt{2}\over a} \left[ 1 - {2\eta \over 3a^2} \right] \,. 
\ee
Alternately, the maximum binary separation for which there 
is a stable equilibrium configuration is given by 
\be
a < a_{max} \approx { \sqrt{2} \over \selfg} \, .  
\label{abound} 
\ee
Note that for a polytropic star, the self-gravity parameter 
$\selfg$ is given by 
\be
\alpha_g(n) = {3 \over 2(5-n)} \,,
\ee
where $n$ is the polytropic index. For low mass stars, $n=3/2$,
$\alpha_g=3/7$, and $a_{max}=7\sqrt{2}/3\sim3.3$. The semimajor axis
is written in units of the stellar radius parameter $R$.  For the
equilibrium state, each stellar radius has the value $R_k$ =
$R(m_k/\mtot)^{1/2}=R/\sqrt{2}$, so that the system has a minimum
separation $a_{min}\sim\sqrt{2}/2\approx0.71$ (note that contact
binaries actually allow for somewhat closer orbits). In addition, 
however, the semimajor axis must satisfy the constraint of equation
(\ref{constraint}) in order for the equilibrium state to be a
minimum. This condition implies that $a\ge(6\eta)^{1/2}$. For
pre-main-sequence stars, the dimensionless moment of inertia
$\eta\approx1/5$, so that $a\ge(6/5)^{1/2}\approx1.10$. As a 
result, the full range of allowed binary orbits for which it is 
energetically favorable to have equal mass is given by
\be
1.1 \lta a \lta { \sqrt{2} \over \selfg} \sim 3.3 \, . 
\label{arange} 
\ee
The dimensionless semimajor axis $a$ is given in terms of the length 
scale $R$, so that the stellar radii $R_k=R/\sqrt{2}$ for equal mass
stars. Newly formed stars and young pre-main-sequence stars are
larger than their main sequence counterparts by factors of $3-4$ (for
solar type stars, e.g., \citealt{stahlerpalla}). Taking these results
into account, the upper bound on the semimajor axis in physical units
becomes 
\be
\aphys \lta 16 \rstarms \,,
\label{physlimit} 
\ee
where $\rstarms$ is the main-sequence radius of the star. One should
also keep in mind that binary separations can evolve after the epoch
of formation. As a result, some binaries with nearly equal masses can
form in tight orbits, within the limit of equation (\ref{physlimit}),
but then increase their semimajor axes afterwards.
In addition, the specific values presented here
depend on the stellar mass-radius relation from equation
(\ref{massradius}). However, the exact form of the mass-radius
relation depends on the mass range of the stars and the stage of
evolution when the stellar masses are determined, so that the limit of
equation (\ref{physlimit}) should be considered approximate (see
Appendix \ref{sec:genforms} for further discussion of how variations
in the stellar structure parameters affect the optimization
procedure).

For completeness, we note that when binaries are close enough to favor
equal mass members, they are also close enough to strongly interact
through magnetic fields during their pre-main-sequence phase. More
specifically, the magnetic truncation radius (e.g., \citealt{shu94})
for circumstellar disks is comparable to the separation of equation
(\ref{arange}). In addition, this orbital separation
  $a\sim10R_\ast$ is roughly the maximum semimajor axis for which
  binary stars are affected by tidal interactions over their main
  sequence lifetimes (e.g., \citealt{hurley}). 

\section{Comparison with Observations} 
\label{sec:obs} 

The analysis of this paper indicates that equal mass stars are
energetically favored, but the critical point is a true minimum only
for close binaries that satisfy the constraint of equation
(\ref{physlimit}).  As outlined in Section \ref{sec:intro},
observations suggest that the population of close binaries displays a
modest excess of nearly equal mass stars, but the distribution of mass
ratios spans the full range of possible values. Nonetheless,
sufficiently close binaries should favor equal mass members. The goal
of this section is to search for such a signature in the currently
available data.

As one example, we consider the sample of spectroscopic binaries
compiled by \cite{mazeh2003}. The paper reports the mass ratios for 
52 single-lined (Table 1) and 11 double-lined spectroscopic binaries 
(Table 2) with primary masses in the range $m_1=0.6-0.85M_\odot$. 
For the single-lined category, observations measure the composite 
quantity $f_M$ given by 
\be
f_M = { (m_2 \sin i)^3 \over (m_1 + m_2)^2} = 
{ (q \sin i)^3 \over (1 + q)^2} \,,
\label{massfun} 
\ee
where $m_1$ ($m_2$) is the mass of the primary (secondary), $i$ is the
inclination angle, and $q$ is the mass ratio.  The mass of the primary
can be independently estimated, so that the observations determine the
minimum value of the mass ratio, $q_{min}$, defined as the value of
$q$ that corresponds to an edge-on orbit ($i=\pi/2$). In contrast,
for double-lined spectroscopic binaries, the additional data available
allow for an estimate of the inclination angle and hence $q$ itself
(see also \citealt{shahaf} for a detailed discussion of the
observational complexities).

\begin{figure} 
\centerline{ \includegraphics[width=0.80\textwidth]{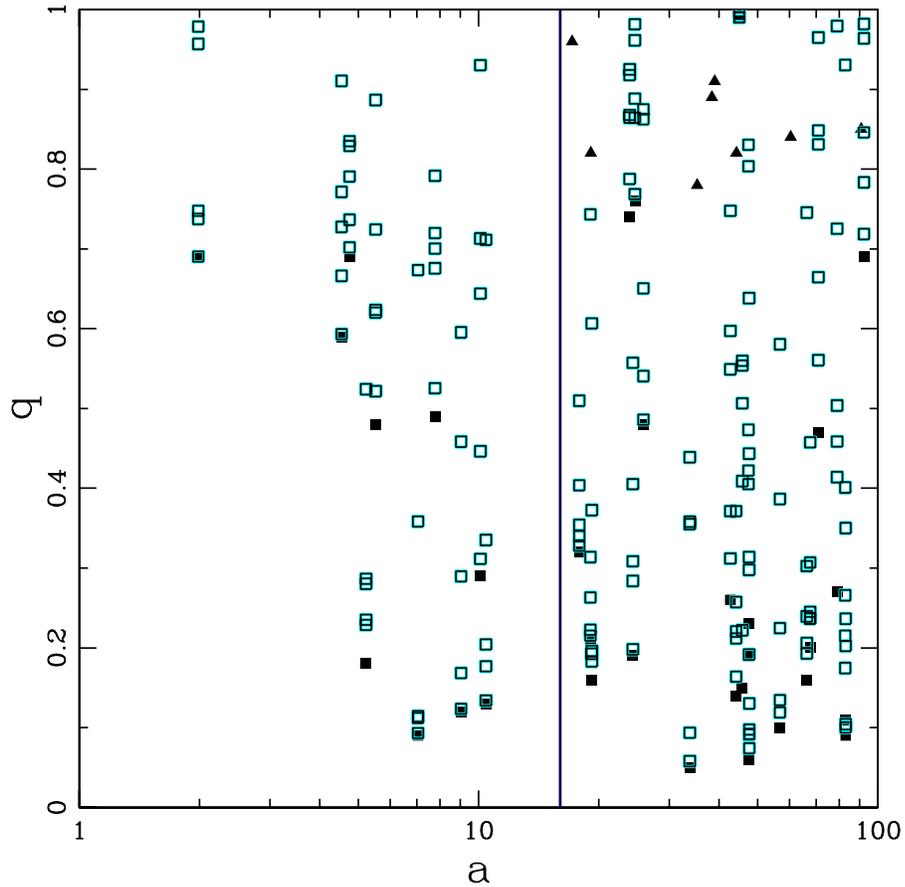} }
\vskip-1.25truein
\caption{Mass ratios $q$ for a collection of spectroscopic binaries  
(from \citealt{mazeh2003}) as a function of separation (in units 
of the primary radius). Solid black squares depict the minimum values 
of the mass ratio $q$ for single-lined spectroscopic binaries. The 
open cyan squares show realizations of the mass ratios obtained by 
sampling the possible inclination angles. The solid black triangles 
depict the mass ratios obtained for the double-lined spectroscopic 
binaries in the sample. The vertical blue line delineates the 
boundary between close binary systems for which equal mass stars 
are favored (left) and wider binary systems (right). } 
\label{fig:obscomp} 
\end{figure}

The observational data from the aforementioned sample are presented in
Figure \ref{fig:obscomp}, which plots the estimated mass ratios versus
the semimajor axes of the systems. Three types of data are presented:
The solid squares show the minimum value of the mass ratio for the
observed single-lined spectroscopic binaries. The open squares show
realizations of the mass ratio obtained from sampling $\sin i$ (and
inverting equation [\ref{massfun}] to find $q$). The mass ratios for
the double-lined spectroscopic binaries are shown as the solid
triangles in the figure. The constraint from equation
(\ref{physlimit}) is shown as the vertical blue line in the diagram.
Although the data span a wide distribution, the points to the left of
the line (for close binaries) tend to have somewhat larger mass ratios
than those on the right (wider binaries), roughly consistent with
theoretical expectations.

For sufficiently close binaries, it is energetically favorable for 
the stars to have equal masses. For binaries wider than the limit
(\ref{physlimit}), which comprise the majority of systems (e.g.,
\citealt{duchene}), it is energetically favorable for one star to
accrete most of the mass. However, the collection of mass ratios shown
in Figure \ref{fig:obscomp} does not show a sharp transition between
close binaries with equal masses and wider binaries with much smaller
mass ratios. In general, observations of binary mass ratios show both
a moderate preference for equal masses and an overall distribution
that is relatively flat (see, e.g.,
\citealt{mazeh1992,duqmayor,tokovinin2000,halbwachs2003}).  The
preference for equal mass stars is often characterized in terms of an
excess of systems with high mass ratios $0.95<q<1$, where the excess
fraction is typically ${\cal F}_{eq}\sim0.1-0.2$ for close binaries 
and decreases significantly with orbital period \citep{raghavan,
moestefano}.  However, the distribution is not skewed heavily toward
extreme values. The fact that this dominance does not usually occur
places constraints on the star formation process. In particular,
binary formation is complicated and energy dissipation is inefficient.
Tidal interactions provide one important source of dissipation for
binary systems, but they have a steep dependence on semimajor axis
\citep{hut1981} and become ineffective for binaries wider than the
limit of equation (\ref{physlimit}).

For binaries that are wide enough so that the equal mass state
($f=1/2$) is not a minimum, we can derive a lower limit on the mass
fraction $f$ (equivalently, the mass ratio $q=f/[1-f]$) by requiring
that the tidal equilibrium state remains a minimum for arbitrary mass
ratios. If we redo the analysis of the previous section, but consider
the mass fraction to be an input parameter rather than a variable to
be optimized over, the criterion (\ref{h3s}) for the tidal equilibrium
state to be a minimum takes the more general form 
\be
\sqrt{a} f (1-f) > 3 \eta a^{-3/2} \left[ f^2 + (1-f)^2 \right] \,.
\ee 
This result implies that the orbital angular momentum must be
greater than three times the spin angular momentum (as found
previously; \citealt{hut1980}). This expression can be rewritten 
as a constraint on the mass fraction 
\be 
{1 \over 2} - {1\over2} 
\left( {1 - 6\eta/a^2 \over 1 + 6\eta/a^2} \right)^{1/2} 
< f < {1 \over 2} + {1\over2} 
\left( {1 - 6\eta/a^2 \over 1 + 6\eta/a^2} \right)^{1/2} \,.
\label{fminimum} 
\ee
In the limit $\eta \ll a^2$, this constraint simplifies to the 
(leading order) form 
\be
{3 \eta \over a^2} < f < 1 - {3 \eta \over a^2} \,.
\ee
The tidal equilibrium state is no longer a minimum of the energy for
mass fractions outside this range (equivalently, for mass ratios 
$q < 3\eta/a^2$). However, this constraint is rather weak: If we
ignore factors of order unity (e.g., $3\eta$) the limit on the
companion mass takes the form $m_2 \gta \mtot (R_\ast/\aphys)^2$. For
solar-type stars, e.g., the limit on $m_2$ falls below the minimum
stellar mass for $\aphys \gta 0.1$ AU.

\section{Conclusion} 
\label{sec:conclude} 

This paper has found the tidal equilibrium states for binary star
systems, including the optimization of the masses of the constituent
members.  In this formulation of the problem, the spins of both stars
and the binary orbit contribute to the angular momentum, which is held
constant. This work generalizes previous treatments by allowing mass
to be apportioned between the two stars, subject to conservation of
total mass, and by including the self-gravity of the stars in the
energy budget.

\subsection{Summary of Results} 
\label{sec:summary}  

The properties of the tidal equilibrium state (equations
[\ref{critzero}, \ref{critpoint}]) are analogous to those found earlier
without considering mass as a variable \citep{hut1980}. The stellar
spins are synchronous and aligned with the binary orbit, which has
zero eccentricity. The existence of the minimum energy state requires
the system to have a minimum total angular momentum (equation
[\ref{lmin}]). In addition, in order for the equilibrium state to be
an energy minimum, the orbital angular momentum must at least three
times larger than the spin angular momentum (equation [\ref{h3s}]).

The main result of this analysis is that the tidal equilibrium state
for binary systems corresponds to equal mass stars. If we include the
self-gravity of the stars in the energy budget, then in order for the
tidal equilibrium state to be a minimum, the semi-major axis of the
binary must be bounded from above (equations [\ref{abound},
\ref{physlimit}]).  As a result, equal mass stars are only
energetically advantageous in sufficiently close binary systems
(equation [\ref{arange}]). The maximum semimajor axis for binaries
that favor equal mass stars occurs for separations $\aphys\sim16\rstarms$
(where $\rstarms$ is the main sequence radius of the star). These
properties are roughly consistent with the observed population of
binary systems, where close binaries have a preference for equal mass
members and wider binaries display a different (more diverse)
distribution of mass ratios.

The treatment of this paper assumes that the mass-radius relation
(\ref{massradius}) holds for both stellar members of the binary.  
It also assumes that both stars have the same structure, i.e., the
constants $\selfg$ and $\eta$ that depend on the internal mass
distribution of the stars are the same for both objects. The finding
that equal mass stars provides the lowest energy state is largely
independent of the particular functional form of the mass-radius
relation, but does depend on the equality of the structure constants
(see Appendix \ref{sec:genforms}). However, the form of the
  mass-radius relation does affect the range of binary separations
  over which the equilibrium state is an energy minimum.  Since the
internal structure of forming and newly formed (pre-main-sequence)
stars evolves with time, the assumption of a common mass-radius
relation requires the stars to form at approximately the same time.

\subsection{Discussion} 
\label{sec:discuss} 

This analysis indicates that energy considerations favor equal mass
stars for sufficiently close binaries, in agreement with observations
(although the picture is complex, as discussed in Sections
\ref{sec:intro} and \ref{sec:obs}). However, most binaries are too
wide for this result to apply, so that the tidal equilibrium state
(with the mass ratio as a variable) becomes a saddle point over most
of the parameter space. Only sufficiently close binaries are affected
by the energy optimization procedure carried out in this paper
(equation [\ref{physlimit}]).

The results presented here are based on global energy considerations,
and are independent of the possible pathways by which the binary
systems realize such states. Nonetheless, some mechanism(s) for
  energy loss must be in operation, and the dissipation time scale
  must be short compared to the time scale for star formation, i.e.,
$t_{\rm diss}\ll t_{\rm form}$.  Although equal mass stars are favored
for close binaries, not all systems will achieve their optimal
state. As a result, close binaries are expected to display a range of
mass ratios, but still show a tendency for mass ratios closer to
unity. For wider binaries, it is not energetically favorable to have
equal masses.  Moreover, the stars in such systems are generally too
far apart to interact, either by mass transfer during the formative
stages or by tidal interactions afterwards, so that the stars can be
considered largely independent.

The constraints of this paper are robust, in that the lowest energy
states are independent of the evolutionary trajectory by which the
binary systems are formed. In particular, the tidal equilibrium state
does not depend on the mechanism(s) by which the system loses energy.
At the same time, this treatment is necessarily incomplete.  Any
source of dissipation can lead to the systems evolving toward lower
energy states, but we would nonetheless like to know the specific
routes through which binary star systems determine their mass ratios
and orbital properties (e.g., see \citealt{tohline,duchene,moekratter},
and references therein). This goal remains a formidable challenge for
the future. 

\medskip 
\textbf{Acknowledgments:} We would like to thank Kaitlin Kratter,
Darryl Seligman, and Chris Spalding for useful discussions. We also
thank the referee for useful input. This work was supported through
the University of Michigan, the National Science Foundation, the Air
Force Office of Scientific Research, the David and Lucile Packard
Foundation, and the Alfred P. Sloan Foundation.

\appendix 
\section{Generalized Forms for the Stellar Structure Parameters} 
\label{sec:genforms} 

In order to carry out the optimization calculation, we had to specify
the mass-radius relationship (\ref{massradius}) and the dimensionless
structure constants ($\eta$ and $\selfg$) for the stars. In this
Appendix we assess how these choices affect the results.  We can
generalize the calculation further in two directions, by considering
the mass-radius relationship to be an arbitrary function, and by
considering the structure constants to have different values for the
two stars.

For a generalized version of the mass-radius relation, the stellar
radii $R_k$ can be written in the form 
\be
R_1 = R \mrs \rof \qquad {\rm and} \qquad R_2 = R \mrs\rofone \,,
\label{genrad} 
\ee
where $\mrs$ is an arbitrary (dimensionless) function and $R$ is 
the same radial scale used previously [so that $\mrs(1)=1$ and 
$R=R_\ast(\mtot)$]. In equation (\ref{genrad}), the function $\mrs$
is evaluated at $f$ for the first star and at $1-f$ for the second
star. We use a smaller font for the arguments to distinguish the
evaluation of the function from the multiplication of the function by
the argument. With this ansatz, the energy of self-gravity becomes 
\be
E_{\rm grav} = - \left[ \alpha_1 {f^2 \over \mrs\rof} + \alpha_2
{(1-f)^2 \over \mrs\rofone} \right] \equiv - 
\left[ \alpha_1 \Gamma\rof + \alpha_2 \Gamma\rofone \right] \,,
\label{gengravity} 
\ee
where the second equality defines a new dimensionless function
$\Gamma$. Note that the $\alpha_k$ can be different for the two stars.
Similarly, the moments of inertia have the generalized forms 
\be 
I_1 = \eta_1 \mtot R^2 \mrs^2\rof \qquad {\rm and} \qquad 
I_2 = \eta_2 \mtot R^2 \mrs^2\rofone \,.
\ee
The dimensionless energy and angular momentum terms for stellar spin
take the modified forms 
\be
E_{\rm spin} = {1\over2} \eta_1 \omega_1^2 \mrs^2\rof + 
{1\over2} \eta_2 \omega_2^2 \mrs^2\rofone \,, 
\ee
and 
\be
{\vvv L}_{\rm spin} = \eta_1 \mrs^2\rof \, {\vvv \omega}_1 + 
\eta_2 \mrs^2\rofone \, {\vvv \omega}_2 \,. 
\ee
For taking the first variation (Section \ref{sec:firstvary}), 
only the derivatives with respect to the mass fraction $f$ are 
affected by the generalization. These new terms have forms 
\be
{\partial E_{\rm grav} \over \partial f} = \alpha_1 
\Gamma^\prime\rof - \alpha_2 \Gamma^\prime\rofone \,,
\ee
\be
{\partial E_{\rm spin} \over \partial f} = 
\eta_1 \omega_1^2 \mrs\rof \mrs^\prime\rof - 
\eta_2 \omega_2^2 \mrs\rofone \mrs^\prime\rofone \,,
\ee
and 
\be
{\partial {\vvv L}_{\rm spin} \over \partial f} = 
\eta_1 \mrs\rof \mrs^\prime\rof \, {\vvv \omega}_1 - 
\eta_2 \mrs\rofone \mrs^\prime\rofone \, {\vvv \omega}_2 \,,
\ee
where the primes represent derivatives of the functions $\Gamma$ and
$\mrs$ with respect to their argument. Given that the critical point
corresponds to ${\vvv \omega}_1={\vvv \omega}_2=\omega{\hat z}$, all
of the above expressions vanish for the choice $f=1-f=1/2$ when the 
structure constants are equal ($\alpha_1=\alpha_2=\selfg$ and
$\eta_1=\eta_2$). As a result, the critical point remains the same,
even for an arbitrary mass-radius relation, i.e., the tidal equilibrium
point corresponds to equal mass stars. 

This result --- equal mass stars --- requires that the functions
specifying the self-gravity and the moment of inertia are the same
for both stars. If they are different functions, either with different
constants ($\alpha_k$, $\eta_k$) or different functional forms
($\Gamma$, $\mrs$), then the tidal equilibrium point will not always
correspond to equal mass stars. If the structure functions are
`almost' the same, then we expect the critical point to have almost
equal mass stars. However, the departure from equality will depend on
the functions in question, so that a more general result is beyond the
scope of this paper.

Similarly, the Hessian matrix elements for the second variation
(Section \ref{sec:secondvary}) depend on the form of the generalized
mass-radius relation $\mrs$ and the ancillary function $\Gamma$. The
range of parameter space for which the eigenvalues are positive, so
that the critical point is an energy minimum, will thus depend on the
form of these functions. For example, if we change the power-law index
of the mass-radius relation of equation (\ref{massradius}), then the
dimensionless function appearing in the generalized self-gravity
contribution of equation (\ref{gengravity}) will have a power-law form
$\Gamma(x)=x^p$. The eigenvalue corresponding to the mass fraction has
the form 
\be
\lambda_f = E_{f\!f} = {3\over a} - {2\eta\over a^3} - 
\selfg\,p\,(p-1)\,2^{3-p}\,,
\ee
where we have assumed that $\alpha_1=\alpha_2=\selfg$. Different
values for the index $p$ will thus change the range of semimajor 
axes for which the eigenvalue is positive and for which the tidal
equilibrium state is an energy minimum.

\bigskip 

\end{document}